# Mental Stress: Source of Neurological Degeneration; Case of MS

**Jahan N Schad***

*Lawrence Berkeley National Laboratory (LBNL), University of California, USA*

**\*Corresponding author:** Jahan N Schad, Lawrence Berkeley National Laboratory (LBNL), University of California, 376 Tharp Drive, Moraga, CA 94556, USA, Tel: 925-376-4126; Email: Jaschadn@gmail.com

**Abstract**

Mental stress is a vague though familiar concept that accounts for an agency of the good, the bad and the ugly, that operates in the brains of animated beings, in response to environmental stressors; a life saver of various degrees, at its best, and indicative of much mental anguish and the likely cause of costly physiological and mental adversities, at its worst. In this work, we provide evidence for correlation of stressors with afflictions of MS throughout the world, and put forward arguments in support of the fact that stressors can render disruptions in the normal computational processes of the brain, defining the innards of the mental stress, which in turn may lead to the onset of physiologic adversities in the biologic systems, possibly rendering various diseases, even one as MS. It can be also deduced that feeling of unease, normally referred to as stress in common usage, is an indicator of a brain process disruption event. This work takes note of the fact that debilitating symptoms and abnormal behaviors have been considered a neurological disorder since around 15th century [1]; and gradual efforts aimed at understanding the phenomena culminated in 1868 work of Charcot [2]: He wrote a complete description of the disease now known as MS- and the changes in the brain that accompany it, based on what had neurologically become known, at the time and his own clinical studies and observations. Much work has been done since then to find the cause of the disease; some of its intermediary process has been recognized and MS is established today as an autoimmune disease, which may also include some characteristics of viral infection as well. While the real cause of the disease is still not known, major focus is put on the treatment of MS symptoms, aiming to slow down its progress and to reduce the frequency of attacks. In this effort we establish the link between MS and mental stress, through analyses of various aspects of statistics of prevalence and incidence, available in the literature [3-5], which lends itself to opening up of additional treatment possibilities that could be used separate of, or conjunctively with, the medical approaches. On a grander scale, publicizing the adverse workings of the mental stress and its evils can attain statistical gains, in the incidence reduction.

**Keywords**

MS; Neurological disorder; Mental Stress; Stressors; Autoimmune System

## Abbreviations

NGS**:** Neuronal Group Selection; AI: Artificial Intelligence

## Background

Symptoms such as, loosing sensation in parts of body, loss of balance, neuritis, tremors, double vision, urinary problems, gets the sufferer to Dr's office and generally in case of severity of any of the symptoms of MRI, Blood test, neurological tests are conducted. Depending on the results of the tests and history of patient, diagnosis of MS, according to certain criteria- based on consensus between Neurologists- may be made. So far, medical research in spite all efforts has remained in the dark regarding the causality aspects of this disease; it is considered an autoimmune, or immune mediated disorder, where brain is targeted, and for which no cure has been found. A number of approaches are available to decrease the frequency and intensity of recurrence of symptoms; however the efficacy and side effects have been of concerns. Prevalence of the MS, costs of medications and care of patients, and the misery of the sufferers, throughout the world, has given it the appearance of a very difficult epidemic type problem, though, almost all are not believed to be of pathogenic origin; and the fact that it affects young lives in their most productive period of life, has made it a worldwide concern. The prevalence and increasing rates of the disease (incidence), has put it on the WHO agenda, and a comprehensive Atlas report [3], by this organization, demonstrates vividly its seriousness. Following excerpt from Pages 14 and 16 of this report, which addresses prevalence and incidence, in total numbers and average age of onset and male/ female ratios, throughout the world, speaks to this fact:

**Salient findings (of total numbers)**

i. Globally, the median estimated prevalenc**e** of MS is 30 per 100 000 (with a range of 5-80).

ii. Regionally, the median estimated prevalence of MS is greatest in Europe (80 per 100 000), followed by the Eastern Mediterranean (14.9), the Americas (8.3), the Western Pacific (5), South-East Asia (2.8) and Africa (0.3).

iii. By income category, the median estimated prevalence of MS is greatest in high-income countries (89 per 100 000), followed by upper middle (32), lower middle (10) and low income countries (0.5).

iv. The countries reporting the highest estimated prevalence of MS include Hungary (176 per 100 000), Slovenia (150), Germany (149), United States of America (135), Canada (132.5), Czech Republic (130), Norway (125), Denmark (122), Poland (120) and Cyprus (110).

v. Globally, the median estimated incidence of MS is 2.5 per100 000 (with a range of 1.1-4).

vi. Regionally, the median estimated incidence of MS is greatest in Europe (3.8 per 100 000), followed by the Eastern Mediterranean (2), the Americas (1.5), the Western Pacific (0.9) and Africa (0.1). No countries in South-East Asia provided data.

vii. By income category, the median estimated incidence of MS is greatest in high-income countries (3.6 per 100000), followed by upper middle (2.2), lower middle (1.1) and low income countries (0.1).

viii. The countries reporting the highest estimated incidence of MS include Croatia (29 per 100 000), Iceland (10), Hungary (9.8), Slovakia (7.5), Costa Rica (7.5), United Kingdom (6), Lithuania (6), Denmark (5.9), Norway (5.5) and Switzerland (5).

ix. The total estimated number of people diagnosed with MS, reported by the countries that responded, is 1 315 579 (approximately 1.3 million) of whom approximately 630 000 are in Europe, 520 000 in the Americas, 66 000 in the Eastern Mediterranean, 56 000 in the Western Pacific, 31 500 in South-East Asia and 11 000in Africa. The reader should keep in mind that there are no data for some of the mega countries such as Russian Federation, where the total number of people has been suggested to be quite high in anecdotal reports.

**Salient Findings (of age of onset and male/ female ratios)**

i. Globally, the inter quartile range for age of onset of MS symptoms is between 25.3 and 31.8 years with an average age of onset of 29.2 years.

ii. Regionally, the average age of onset is lowest in the Eastern Mediterranean (26.9) followed by similar average age of onset in Europe (29.2), Africa (29.3), the Americas (29.4), and South-East Asia (29.5) and highest in Western Pacific (33.3).

iii. By income category, the estimated average age of onset is 28.9 years for the low and upper middle-income countries and 29.5 and 29.3 years for high and lower middle-income countries

iv. Globally, the median estimated male/ female ratio is 0.5, or 2 women for every 1 man (with a range of 0.40 to 0.67).

v. Regionally, the median estimated male/ female ratio is lowest in Europe (0.6), the Eastern Mediterranean (0.55) and the Americas (0.5) and highest in South-East Asia (0.4), Africa (0.33) and the Western Pacific (0.31).

vi. By income category, the median estimated male/ female ratio is same in all income groups of countries (0.50).

Above data contains shocking information about the nature of distribution of the disease that could be summarized in the following distinct facts:

a) Low age of onset, 20 to 35.

b) Similar age of onset age among all income groups.

c) Low prevalence around Equator, mostly less developed as well.

d) High prevalence in industrial countries.

e) High prevalence among higher-income groups.

f) High prevalence among educated groups.

g) Female/ male ratio of 2, for MS worldwide, increasing to 3, or more, in less developed regions.

Also, the followings are anecdotal extract from the report, which in one instance draws upon information from additional source:

a. High prevalence and incidence in harsh political and economic conditions in some third world countries- Example of Iran: Incidence rate has climbed drastically in about 30 years- since 1979 revolution- resulting in change of prevalence from around 10 to around 60 per 100,000, making it close to those of the industrial nations and affection age groups average of 25 or less [4].

b. Increasing rates in low incidence countries with recent economic growth [3].

c. Recognized possible role of one or more combination of environmental factors [1].

Added to the panoply of data, we need to also bring in the following medically observed facts about the disease, which speak to its main characteristic:

- Varying frequency of attacks.
- Varying intensity of attacks.
- Varying nature of symptoms.
- Occasional lifelong dormancy of attacks.

Having provided what is mainly known of the distribution of the disease, and some the well-known facts, we now proceed to provide the context: in which our arguments are pursued.

**Introduction**

Autoimmune system attacks, in the brain, are the medically recognized causes of a number of physiological disorders and diseases. Given this fact, the assumption of occurrences of antecedent events in the brain, serving as triggers for the attacks, is not unreasonable. Since no symptoms seemingly exist prior to the attacks, the a priori events have to be non-perceptible and presumably not involving any measurable physical deterioration of the brain; very likely some non-physical disruption*s* in the normal processes of the brain. This, in the context of today understands of the brain, from the viewpoints of the fields of philosophy, psychology, cognition, linguistics, and neurosciences, has to be considered a disruption in its normal computational processes. The thought of computational brain, which (seemingly) seriously originated from Descartes ponderings about the clockworks automata, has engendered many efforts in understanding of brain processes through research in computational neurosciences, resulting in theories such as Neuronal Group Selection (NGS), and Reductionist and Connectionist Artificial Intelligence (AI), altogether defining new the science of the mind: "This new science of mind is based on the principle that our mind and our brain are inseparable. The brain is a complex biological organ possessing immense computational capability: it constructs our sensory experience, regulates our thoughts and emotions, and controls our actions. It is responsible not only for relatively simple motor behaviors like running and eating, but also for complex acts that we consider quintessentially human, like thinking, speaking and creating works of art. Looked at from this perspective, our mind is a set of operations carried out by our brain. The same principle of unity applies to mental disorders by Noble Laurette Eric Kandel [6].

From computational angle, the (brain) biology inspired scientific neuronal and neural networks- both created with underlying assumptions of non symbolic and non logic driven brain processes- that have demonstrated capabilities in solving complex problems, with no need for logical formalisms, unlike traditional digital computers that depend on it, are only glimpses into the brain's computational power, and what is known about it, though substantial, is still very far from complete. However, Brain, this amazing biological computational system, like its (comparatively much simpler) counterparts in the physical world, is subject to both attacks and malfunctions; though, unlike them, it does not suffer from regular down-time, the quality of its outputs can be adversely affected. This depends on the efficacy of its in-built redundancy and on-the-fly biological repairs, in averting damage. The hypothesis set forth here, is addressing one major instance of attack: by diagnosing its mechanism, implementation of possible preventive, or alternative treatment, measures may become possible.

## Hypothesis

The correlations of the MS afflictions with some of the seemingly advantages elements of life, in the industrial and modern era, are indicative of (mostly unconscious) heavy demand on the minds, which such conditions inherently begets. Likewise, the correlations of harsh, societal, economical, and environmental conditions- generally politically induced in today's world- with the prevalence of the disease, again point at the prevailing heavy mental engagement as a result of copying with these adverse life-affecting elements. It is therefore clear that high incidence and prevalence of the disease are symptomatic of excessive engagements of the mind with irresolvable complexities posed by the modern life environmental stressors, possibly rendering disruptions in brain's normal operations. In the context of a computational model for the (mental) functional operations of the brain, any non-physical disruptions in the brain processes immediately indicates its engagement with irresolvable computational problems, that can adversely bears on the physiological processes which are computationally controlled by the brain. Early indication of this dilemma is perceptions (feelings) of measures of unease, commonly referred to as stress.

Stressors have been a fact life in any era of human civilization, however since stressors of different nature and magnitude have been added in the post industrialization era, their nature and implications would be different from those of the past; with the advent of the industrial age, progress, as well as problems, began affecting human life very dramatically, mostly for the better and at times for the worse. One of the major adverse characteristics of the new era is the presence of high "sustained" stressor levels though mostly directly unnoticed-, a condition that very likely would not develop in the prior simpler life on the planet. Obviously, human beings suffered much from many hardships in the pre-industrialization era. However, those mostly related to the issues of food and shelter; adverse attributes of thoughts and desires of much material ownership, and other worldly accolades, since mostly limited, did not have significant contributions to it. Of course wars and pestilence (not affecting major segments of the humanity today, yet!), would bring occasional very high peaks of stressors. But the resulting stresses, whether they were long lasting or not, were handled accordingly through mental abilities, gained in the process of natural living. Today's high levels of the (background) "sustained" mental stresses- mostly not (natural) survival driven- are brought about by new stressors: challenges of modern life, goals of high professional and material achievements, as well as, by the (mental) demanding efforts of chasing the (mainly) illusory premises of materialistic or fantasy based happiness, in the postindustrial world; drastic, changes and changing of habitat, the modern life styles, increasing family instability, along with the many synthetic compounds in the environment, all mostly unfamiliar to human biology and psyche are also significant contributors to the experienced high stress levels, as well as, causes of health problems, though some of the former can also be consequents of the latter.

As mentioned earlier, stressors are discerned as complex problems in the brain and when they are irresolvable, their effects, based on scientific inference from the present understanding of brain functions, must be hampering of proper deployment of neural patterns (Brain web constructs of organized and synchronous neural signal firing schemes) that control most inner body functions, and its activities. This may drastically affect hormonal balances, as well as, other bodily functions, upsetting homeostasis, the equilibrium mainly regulated by brain signals received through the nervous system, engaging the Endocrine systems, the Autonomic nervous system, and the Immune system as needed. The state of the brain that leads to any measure o*f* disequilibria of the biologic system is a stressed state and its immediate effects on parts of the Limbic system, may create abnormal sensations, perceptions of which are what commonly are referred to *a* stress, as indicated earlier. Species survival have obviously depended on certain level of stress, that is, stressor originated effect for certain necessary physiologic and bodily responses, such as fight or flight and other survival benefits; however, sustained levels of high stress, that is continued disruption of brain functional processes, in view of links with the rest of the biological system, may initiate upheavals in the body, forcing the immune system to rise to the occasion, as is the case in MS, attacking whatever and wherever the source of the problem is; in this case, it attacks the brain to eradicate the cause; *a* response expected from a "strong" defense system! Another reason for attack of the immune system would be presences of infectious viruses that can act as a trigger, however, since prolonged presence of viruses are not likely, this scenario does not conform with observed long term persistence of MS; perhaps whatever short term neurological effects emanate from the Viral infections could be put in the category of the results of known MS afflictions that have been found to be of very limited consequence in life. In MS, the attack removes myelin from a number of the nerve cells (affecting the quality of the neural signals, at least temporarily), which may get repaired, perhaps immediately. Mild attacks affecting small areas of the brain cells may go unnoticed. Repeated autoimmune attack leaves some cells with lesions. If such attacks continue more severely, and/ or for long time, more parts of the brain may be affected and very dramatic events in the body may result. Field evidence suggests these attacks may, in some cases, stop at some point and never happen again, however, in other cases there are stop-and-go episodes leading to various stage of the progression of the disease.

Considering the high prevalence and incidence of MS, in the stable advanced industrial countries, as well as in the rapidly progressing (developing) countries, on the one hand, and in parts of the developing world with socioeconomic and political instability, on the other hand affecting women much more than in the former cases-, the commensurate stresses, the obvious indicators of life in such high stressor environments, would very likely play major role in the onset of the disease; indicative of strong correlation with both modernity and modern misery- the latter applies worldwide.

The two exceptions of the industrial countries, Netherlands and Israel, with much smaller prevalence can perhaps be explained with presence of very strong social welfare systems, and cultural ties in the population; both have major role in stress reduction. The lower statistics in less developed societies, though varied, may be attributed to the prevalence of lower, or at best normal, strength immune systems evidenced by occurrences of all kinds of bodily malfunctions, among their population. Somewhat of anecdotal evidence may be in the fact that people more affected by infectious diseases (possibly due to lower strength of immune system), are less likely to get MS. of course lesser pressures of modern life would be an added factor, as well.

**Tests of the hypothesis**

Stress management [7] has already proven its effectiveness, in dramatic reduction of attacks in patients; it serves as an initial step in the path to confirmation of the hypothesis. Important evidence that has favorable bearing on the hypothesis is the widely accepted curative role of Placebo effect; its efficacy is proven in many clinical trials where hopes of a cure fight off the effects of certain stressors.

Validity of the hypothesis can be investigated in the likes of the following studies:

1) Development and implementation of vigorous stress management program on groups of patients whose attacks happened in highly stressed period of their lives, and monitoring of the results.

2) Evaluation of episodic attack in patients to find out if they coincide with high stress period, and monitoring the rate of the creeping cell deterioration, during the quiet period.

3) Investigation of conditions of the lives of all once afflicted people who have lived a normal life to the old age, finding out when and where they were afflicted, in order to find the needed correlation (reportedly some 20% of diagnosed patients are of such fortune)

4) Collection and analyses of fMRI results that is done unrelated to MS, in elderly people, addressing any presence of Lesions and with no MS symptoms.

**Corollary**

It follows directly or by inference, from the hypothesized genesis of the disease that in the development of brain lesions, stress intensity, strength of the autoimmune system, as well as levels of the agility of the minds, and last but not the least, the mental attitude, matter significantly. Brain agility, problem (stress) resolution ability, life style, as well as positive outlooks in life, are all important factors in avoiding the disruptive actions of stressors; only abnormal out of control circumstance may be over whelming enough to induce serious disruption in brain processing function.

Considering the episodic action of the autoimmune system, in MS, it is plausible that after an attack, relief from all kinds of hanging (sustained) stressor posed problems happens (an electric re-set of sorts)-excepting those of the relevant ordinary concerns. That is, the stress related neural patterns are modified and/ or destroyed and erratic signals are stopped (neural pattern signal weights need to be reconfigured depending on the subsequent conditions of the life and living). Perhaps that leaves the brain with a much cleaner slate, at least for a while, and hence no need for continuation of attacks- the enemy is stopped, at least temporarily. In some afflicted persons (the lucky ones), the levels of pre-attack stress are never reached again- post attack condition of patient's life can perhaps be conducive to permanent automatic removal of high stressors, and therefore stresses- and a hundred percent normal life continue. And, in some other cases, life style changes, and/ or other natural or medical measures, may either prevent continuation of that level of stress, or confuse the immune system preventing future attacks, to allow a normal life to go on. Deploying the plasticity of the brain, to enhance its data processing and complexity resolution capability, through learning and engagement with Arts and Sciences, learning of new skills, would serve as a tremendous added measure to help a post syndrome normal life.

## Conclusion

The work theorizes that the stressor induced mental stress is the reason for the observed direct correlation that is found between environmental stressors and the onset of MS. Stressors stem from adverse aspects of the modern life and times, which are discerned as irresolvable complexities in the brain, hampering its normal computational processes that control the biologic equilibrium of the physiologic system. The hypothesis finds support in stress management studies, and indirectly from Medicine's acknowledged curative role of placebo, which establishes mind's role in disease treatment. As such, the role of stress in any disease, compared to that of pathogenic factors, should be given due weight in diagnosis, as well as in prognosis. This view of scrutinizing life under much higher "sustained" levels of stress a circumstance that the evolutionary path did not prepare us for- carries much substance.

## Acknowledgement

Thanks are due to F. S. Sakha, a researcher in medical treatment of MS, for the discussion we had and for the help with the literature.